\documentclass[letter,traditabstract]{aa} 
\usepackage{graphicx}
\usepackage{txfonts}
\usepackage{natbib}

\bibpunct{(}{)}{;}{a}{}{,} 
\usepackage{mathcomp}
\usepackage{graphicx}
\usepackage{txfonts}
\usepackage{longtable}
\usepackage{lscape}
\usepackage{wasysym}

\renewcommand{\d}{\mathrm{d}}

\newcommand{\Mearth}{$\mathrm{M_\oplus}$}

\newcommand{\Rsun}{\mathrm{R_\odot}}

\newcommand{\etal}{et al.}

\newcommand{\micron}{$\mu$m}
\newcommand{\Rvenus}{\mathrm{R_{\venus}}}
\begin{document}
   \title{Transmission spectrum of Venus as a transiting exoplanet}

   \author{D.~Ehrenreich\inst{1}, A.~Vidal-Madjar\inst{2}, T.~Widemann\inst{3}, G.~Gronoff\inst{4}, P.~Tanga\inst{5}, M.~Barth\'elemy\inst{1}, J.~Lilensten\inst{1}, A.~Lecavelier~des~Etangs\inst{2} \& L.~Arnold\inst{6}}
          
   \institute{
   	UJF-Grenoble 1 / CNRS-INSU, Institut de Plan\'etologie et d'Astrophysique de Grenoble (IPAG) UMR 5274, Grenoble, France
    	\email{david.ehrenreich@obs.ujf-grenoble.fr}
    \and
	Institut d'Astrophysique de Paris, Universit\'e Pierre \& Marie Curie, CNRS UMR 7095, Paris, France
	\and
	LESIA, Observatoire de Paris, CNRS, UPMC, Universit\'e Paris-Diderot, Meudon, France
	\and
	NASA Langley Research Center, Science Directorate, Chemistry and Dynamics Branch, Hampton, Virginia, USA
	\and
	Laboratoire Cassiop\'ee UMR 6202, Universit\'e de Nice Sophia-Antipolis, CNRS, Observatoire de la C\^ote d'Azur, Nice, France
	\and
	Observatoire de Haute-Provence, CNRS/OAMP, Saint-Michel l'Observatoire, France
             }

   \date{}

 
  \abstract{On 5--6 June 2012, Venus will be transiting the Sun for the last time before 2117. This event is an unique opportunity to assess the feasibility of the atmospheric characterisation of Earth-size exoplanets near the habitable zone with the transmission spectroscopy technique and provide an invaluable proxy for the atmosphere of such a planet. In this letter, we provide a theoretical transmission spectrum of the atmosphere of Venus that could be tested with spectroscopic observations during the 2012 transit. This is done using radiative transfer across Venus' atmosphere, with inputs from in-situ missions such as \emph{Venus Express} and theoretical models. The transmission spectrum covers a range of 0.1--5~\micron\ and probes the limb between 70 and 150~km in altitude. It is dominated in UV by carbon dioxide absorption producing a broad transit signal of $\sim 20$~ppm as seen from Earth, and from 0.2 to 2.7~\micron\ by Mie extinction ($\sim 5$~ppm at 0.8~\micron) caused by droplets of sulfuric acid composing an upper haze layer above the main deck of clouds. These features are not expected for a terrestrial exoplanet and could help discriminating an Earth-like habitable world from a cytherean planet.}


   \titlerunning{Venus as a transiting exoplanet}
   \authorrunning{Ehrenreich \etal}

   \maketitle
%

\section{Introduction}
On 5--6 June 2012, Venus will be transiting the Sun as seen from the Earth, for the last time until 2117. This rare astronomical event was previously observed and reported in the literature six times only. The first detection of the atmosphere of Venus (the cytherean atmosphere) is traditionally attributed to Lomonosov, who observed the 26 May 1761 transit from the observatory of Saint-Petersburg \citep{Marov:2005}. Today, modern approaches are used to detect the atmospheres of transiting exoplanets. Several studies using the transmission spectroscopy technique have provided significant insights into the atmospheric composition, structure, and dynamics of hot giant exoplanets \citep[e.g.,][]{Charbonneau:2002,Vidal-Madjar:2003,Vidal-Madjar:2011,Snellen:2010}
This technique is now attempted on Neptune-mass exoplanets \citep{Stevenson:2010,Knutson:2011} and on ``super-earths" \citep[1--10~\Mearth;][]{Bean:2010,Bean:2011,Desert:2011a}. The next step, characterising the atmospheres of Earth-mass planets with transmission spectroscopy, is extremely challenging because of the small spatial extent of these gas envelopes: photometric precisions of the order of 0.1~ppm should be reached for that purpose \citep{Ehrenreich:2006b, Kaltenegger:2009a}. The class of Earth-mass planets includes telluric --\,rocky\,-- planets such as the Earth, Venus, or Corot-7b \citep{Leger:2011} and so-called ``ocean-planets'' \citep{Leger:2004}, possibly such as GJ~1214b \citep{Charbonneau:2009}. Among planets in this mass range, the atmospheres of telluric exoplanets are the most challenging to characterise because these dense planets have small atmospheric scale heights, and thus compact atmospheres \citep{Ehrenreich:2006b}.

In this context, Venus can provide an essential proxy for a telluric exoplanet. Obtaining its transmission spectrum during a transit across the Sun will serve both as a comparison basis for transiting Earth-mass exoplanets to be observed in the future, and a proof of feasibility that such observations can effectively probe the atmospheres of exoplanets in this mass range. In addition, transit observations of Venus can bring precious information about how the atmosphere of a non-habitable world -- observed as an exoplanet -- differ from that of a habitable planet, the Earth, also observed as an exoplanet in transit during Lunar eclipses \citep{Vidal-Madjar:2010}. The previous transit of Venus in 2004 was the first to be scrutinised with modern instrumentation, from space \citep{Schneider:2006a,Pasachoff:2011} and from the ground \citep{Hedelt:2011,Tanga:2011}. This letter aims at providing a theoretical transmission spectrum of the atmosphere of Venus, from the ultraviolet to the infrared, as it could be observed during the transit of June 2012.
 
\section{Model}

\subsection{Atmosphere}
Transmission spectroscopy probes the atmospheric limb of the transiting planet. In the following, we consider an unidimensional atmospheric model, only varying with respect to the altitude and ignoring latitudinal and longitudinal variations in atmospheric structure and composition. Although latitudinal variation observations are well documented (e.g., in the atmospheric density and temperature, cloud top altitude, etc.), this simplification is correct considering the geometry of the experiment and the lack of spatial resolution: the latitudinal variations are averaged in the transmission spectrum of the whole limb.

\begin{figure}
\resizebox{\columnwidth}{!}{\includegraphics{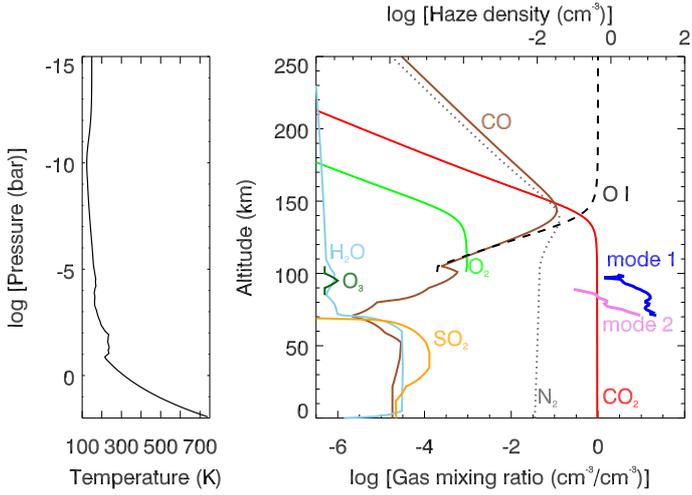}}
\caption{\label{fig:profiles} Atmospheric profiles used in the model. \emph{Left:} pressure-temperature. \emph{Right:} Gas mixing ratios (thin lines) and haze densities for mode-1 (thick violet line) and mode-2 (thick blue line) particles from \emph{Venus Express} data \citep{Wilquet:2009}.\label{fig:profiles}}
\end{figure} 

We consider an altitude range comprising the troposphere (0--60~km), the mesosphere (60--100~km), and the upper atmosphere ($>100$~km) of Venus, up to 400~km in altitude. \citet{Patzold:2007} measure the mesospheric temperature profile $T(z)$ with radio sounding from the VeRa instrument on \emph{Venus Express}. We use their profile assuming an upper boundary temperature of 200~K at 100~km. The tropospheric temperature profile is extrapolated from the tropopause (220~K at 61~km) down to the surface assuming a constant temperature lapse rate of +8.5~K~km$^{-1}$, yielding a surface temperature of 740~K. Above 100~km, the temperature profile corresponds to the neutral component of the ionosphere and follow those used by \citet{Gronoff:2008}, initially calculated by \citet{Hedin:1983}. The pressure profile $p(z)$ is calculated across the whole altitude range (0--400~km) assuming the atmosphere behaves as a perfect gas at hydrostatic equilibrium, $p(z) = p_0 \exp\left[{-\int{{\d z}/{H(z)}}}\right]$, where $p_0 = 93$~bar is the surface pressure and $H(z) = k_B T(z) / [\mu(z) g(z)]$ is the atmospheric scale height, with $g(z)$ the acceleration of gravity, $k_B$ the Boltzmann constant, and $\mu(z)$ the mean atmospheric molar mass. 

\emph{Atoms and molecules\,---}
The mean molar mass variations with altitude are small from the ground up to $\sim120$~km, where carbon dioxide dominates the atmospheric composition (volume mixing ratio $X_\mathrm{CO_2}=96.5\%$). Above this altitude, $X_\mathrm{CO_2}$ follows the model of \citet{Gronoff:2008}. In addition to CO$_2$, we are considering molecular nitrogen ($X_\mathrm{N_2} = 3.5\%$), molecular and atomic oxygen (O$_2$ and \ion{O}{i}), carbon monoxide (CO), water (H$_2$O), and sulfur dioxide (SO$_2$). The mixing ratio profiles for CO, SO$_2$, and H$_2$O are extrapolated from the values from \citet{Cotton:2011}, \citet{Bergh:2006}, and \citet{Titov:2009}, respectively. Mixing ratios of O$_2$ and \ion{O}{i} are considered to be non-zero only in the upper atmosphere, where these species are produced; there, their mixing ratio profiles follow those calculated by \citet{Gronoff:2008}. We finally include mesospheric ozone (O$_3$) following its recent detection by \citet{Montmessin:2011}.

\emph{Clouds and haze\,---}
The thick deck of clouds extending across the troposphere and a part of the mesosphere (45--70~km) is treated as an achromatic optically thick layer. This assumption is justified by the effective Mie scattering from particles with a broad (multimodal) size distribution composing the main deck of clouds \citep{Wilquet:2009}, over the considered wavelength range (0.1--5~\micron). Practically, this means that no signatures can be detected below the top of the cloud deck, set at an altitude of 70~km \citep{Ignatiev:2009,Pasachoff:2011}. We distinguish, however, this main cloud deck from the upper haze (70--90~km). While both are mainly composed of sulfuric acid (H$_2$SO$_4$) aerosol particles, the particle size distribution is different in these layers, yielding different scattering properties. In the upper haze, particles are smaller than in the cloud deck, with typical radii from 0.1 to 0.3~\micron, as expected from pure ``mode-1'' particles \citep{Knollenberg:1980,Esposito:1997}. Following \citet{Wilquet:2009} and references therein, we consider an upper haze consisting of concentrated (75\%) aqueous and spherical droplets of H$_2$SO$_4$. The size distribution of mode-1 droplets is described by a log-normal distribution with parameters $r_g=0.2$~\micron\ and $\sigma_g = 0.2$. \citet{Wilquet:2009} also present the first evidence for a bimodal distribution of aerosol particles in the upper haze of Venus, similar to the two modes in the upper clouds, at the latitudes probed by \emph{Venus Express}: ``mode-2'' particles, with $0.4 \la r_g \la 1.0$~\micron. In the following, we consider that mode-2 particle sizes follow a log-normal distribution with $r_g = 0.6$~\micron\ and $\sigma_g = 1.5$. The vertical density profiles of upper haze mode-1 and mode-2 particles are constrained from data from the UV channel of the SPICAV instrument and the SOIR instrument, respectively, on board \emph{Venus Express} during orbit n.~485 \citep[][ see their Fig.~9]{Wilquet:2009}.

\emph{Radiative transfer\,---}
The radiative transfer model is based on \citet{Ehrenreich:2006b}. It is an unidimensional, single-scattering code calculating the atmospheric opacities along line of sights crossing the limb of Venus. The opacities are then integrated over the whole limb to give a wavelength-dependent effective surface of absorption. The ratio of this surface to the surface of the Sun is the depth of the transit, as measured for exoplanets. In this frame, the transit depth can be converted into a radius ratio and, knowing both the radii of Venus and the Sun, to the effective height $h$ of absorbing species in the cytherean atmosphere, as defined in Sect.~\ref{sec:results}. During the transit on 5--6 June 2012, Venus will be approximately three times closer to the Earth than the Sun. Consequently, the angular diameters of Venus and the Sun will be 57\farcs8 and 31\farcm5\footnote{\texttt{http://ssd.jpl.nasa.gov/horizons.cgi}.}, respectively, and the transit depth measured from Earth will be the ratio between these solid angles, $\delta_\mathrm{\venus} = 934$~ppm. If Venus were a real exoplanet transiting a Sun-like star, the transit depth $\delta$ would be $75.7$~ppm. We thus define a ``geometric'' factor $\gamma = \delta_\mathrm{\venus}/\delta = 12.3$.

\emph{Photoabsorption\,---}
The model computes opacities from photoabsorption cross-sections of the atmospheric components. These cross-sections give rise to the spectroscopic signatures that will be sought for during the transit. Infrared molecular cross-sections are calculated from the HITRAN~2004 line list \citep{Rothman:2005} following \citet[][ Sect.~A.2.4]{Rothman:1998} but using Voigt functions instead of Lorentzian line profiles. The UV cross-sections of CO$_2$, O$_2$, SO$_2$, and O$_3$ are taken from the AMOP\footnote{\texttt{http://amop.space.swri.edu/}} and AMP\footnote{\texttt{http://www.cfa.harvard.edu/amp/ampdata/}} data repositories. The photoabsorption cross sections are plotted in Fig.~\ref{fig:spectrum}a.

\emph{Scattering\,---}\label{sec:scattering}
Important additional absorption is caused by diffusion processes: Rayleigh scattering from atmospheric CO$_2$ and, to a lesser extent, N$_2$, CO, and H$_2$O, emerging above the main cloud deck at 70~km, and Mie scattering caused by H$_2$SO$_4$ droplets in the upper haze (70--90~km). Rayleigh diffusion, which efficiency increases toward the blue following $\lambda^{-4}$, results from the scattering of light by particles with sizes $r$ much smaller than the wavelength ($x = 2\pi r / \lambda << 1$), typically, molecules composing the atmospheric gases. In the case of Venus, the diffusion is mainly caused by CO$_2$: the scattering cross section is $8/3 x^4 (n^2-1)^2 / (n^2+1)^2$ \citep[see, e.g.,][]{Etangs:2008a}, where $n$ is the refractive index of the gas. For CO$_2$, we use the formula of \citet{Sneep:2005} to calculate $n$ as a function of wavelength. The upper haze is composed by particles with sizes larger than the wavelength ($4\la x\la 300$). Consequently, the calculation of the extinction (scattering + absorption) cross section by the bimodal distribution of upper haze particles is based on Mie theory. The complex refractive index of H$_2$SO$_4$ droplets is taken from \citet{Hummel:1988}. Below 3~\micron, the imaginary part of the refractive index is negligible and the scattering dominates the Mie extinction. The absorption cross section, linked to the imaginary part of the refractive index, becomes non-negligible above $\sim3$~\micron, whereas the scattering cross section drops. The extinction cross section is the sum of the scattering and absorption. It is calculated for log-normal particle size distributions with the set of light scattering routines available at the University of Oxford Physics Department\footnote{\texttt{http://www-atm.physics.ox.ac.uk/code/mie/index.html}} and plotted as a function of wavelength for hazes with different particle size distributions in Fig.~\ref{fig:spectrum}a.

\section{Results and discussion}
\label{sec:results}

\begin{figure}
\resizebox{\columnwidth}{!}{\includegraphics{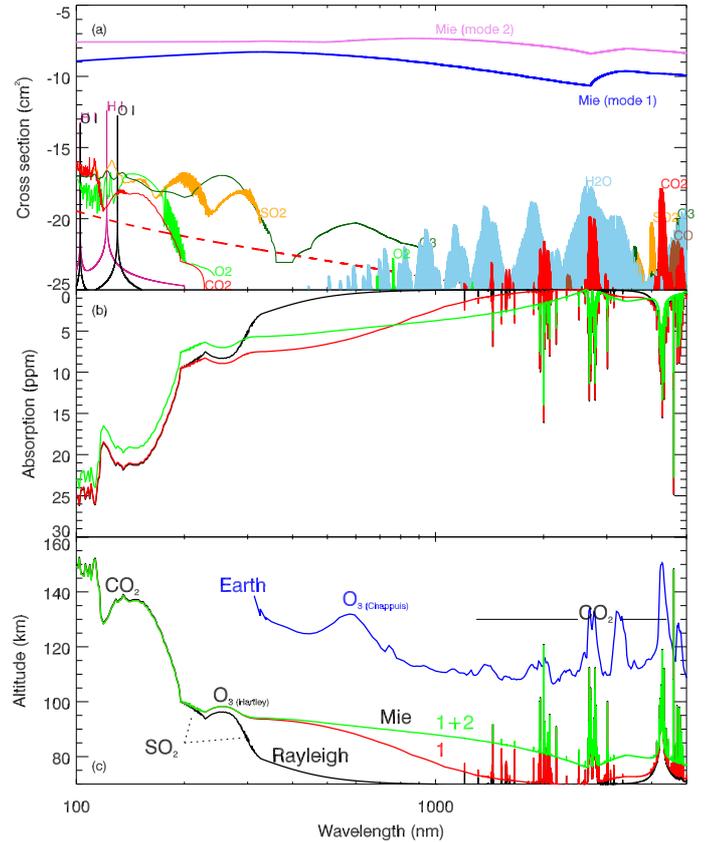}}
\caption{\label{fig:spectrum}\emph{(a)} Absorption cross sections of the considered atmospheric gases, Rayleigh scattering cross section of CO$_2$ (dashed red line), and Mie extinction cross sections for mode-1 (thick blue line) and mode-2 (thick violet line) haze particles. \emph{(b)} Transit spectrum of Venus transiting in front of the Sun, as seen from Earth, in relative absorption and \emph{(c)} effective height of absorption. The different transit spectra shown are models with no upper haze (black), mode-1 upper haze (red), and modes-1+2 upper haze (green). The top of the main cloud deck is set at 70~km. The transit spectrum of the Earth calculated by \citet{Kaltenegger:2009a} is overplotted (blue) and shifted by $+100$~km for clarity.}
\end{figure}

The transmission spectrum of Venus is shown with a resolution of 1~nm in Fig.~\ref{fig:spectrum}b, as relative absorption $\Delta\delta_\mathrm{\venus}(\lambda) = \delta_\mathrm{\venus}(\lambda) - \delta_\mathrm{\venus}(\lambda_\mathrm{min})$, where $\delta_\mathrm{\venus}(\lambda) = \{[\Rvenus + h(\lambda)]/\Rsun\}^2$ and $\lambda_\mathrm{min}$ is defined as the wavelength where the transit depth is minimal; it is 1.67, 2.50, and 2.65~\micron\ for the haze-free, mode-1 haze, and modes-1+2 haze models, respectively. The effective height of absorption $h(\lambda)$ has the same meaning as in \citet{Kaltenegger:2009a} and is plotted in Fig.~\ref{fig:spectrum}c. It can be retrieved on-line (Table~\ref{tab:spectra}). The amplitude of the spectrum reaches $\sim 25$~ppm for the prominent CO$_2$ UV bands below 0.2~\micron. From 0.2~\micron\ and above, the spectrum is dominated by Mie scattering ($<2.7$~\micron) and absorption ($>2.7$~\micron) by upper haze particles, which can significantly impact on the amplitude of the absorption from the other spectral features, such as the CO$_2$ transitions around 2~\micron\ or the Hartley UV band of O$_3$, depending on the assumed particle size distribution. The maximum amplitude of this effect (between UV and IR) is $\sim 6$ to 8~ppm for mode-1 and modes-1+2 haze models, respectively. In the infrared, the most noticeable feature is the $\nu_3$ vibrational band of CO$_2$ at $\sim4.3$~\micron\ ($\sim15$~ppm). For a real exoplanet transit, these absorption would be a factor $12.3$ smaller because there is no parallax effect -- the distance to the exoplanet equals the distance to the transited star.

In terms of effective height of absorption $h(\lambda)$ (Fig.~\ref{fig:spectrum}c), the cytherean limb could be probed from 70 to 150~km, i.e.,\ from the top of the cloud deck up to above the mesopause, in case the upper haze is only composed by mode-1 particles. If larger mode-2 particles  are involved, it won't be possible to probe the limb below $\sim80$~km, reached at 2.65~\micron. 

In fact, the lowest altitude that it is possible to reach with transmission spectroscopy is set by the dominant diffusion regime, Rayleigh or Mie. In the case of Venus, the most remarkable and extended spectral signature is that of Mie scattering by the upper haze. This situation contrasts with the Earth, whose transmission spectrum is dominated at short wavelengths by Rayleigh scattering from N$_2$ and the broad Chappuis band of O$_3$, as modelled by \citet{Ehrenreich:2006b}. This first basic modelling has been refined in the case of the Earth by \citet{Kaltenegger:2009a}, whose spectrum is reproduced in Fig.~\ref{fig:spectrum}c. 

In an atmosphere where Rayleigh scattering is the dominant diffusion process, it is fairly straightforward to assume that, eventhough there is no spectral identification from spectral lines, the main carrier of the Rayleigh scattering is the most abundant atmospheric gas. That would be H$_2$ for a giant exoplanet \citep{Etangs:2008b} or N$_2$ and CO$_2$, for the Earth (or Titan) and Venus, respectively. Because the atmospheric scale height can be calculated from the Rayleigh scattering signature, if the temperature is measured or modelled independently, this signature provides an estimation of the atmospheric mean molecular mass and thus some information on the atmospheric composition. The situation is, however, more complex when Mie scattering is involved because it cannot be simply attributed to the dominant atmospheric species, as  illustrated in the case of the hot jupiter HD~189733b, which transmission spectrum appears dominated by diffusion from the visible to the near-infrared \citep{Pont:2008,Sing:2011a}. \citet{Etangs:2008a} assess the possibility of Mie scattering by a haze layer and propose enstatite particles (MgSiO$_3$) as scattering carriers, since their refractive index is compatible with the observed wavelength range of the Rayleigh signature. In this case, however, the identification is tentative mainly because we have no real proxies of such a planet in the Solar System.

There are a couple of other remarkable molecular signatures besides CO$_2$ ones. In spite of a low mixing ratio ($10^{-7}$ to $10^{-6}$) and a localised vertical distribution \citep[$\sim90$--100~km;][]{Montmessin:2011}, the Hartley band of O$_3$ emerges from the Mie scattering. This somewhat surprising spectral feature is actually enhanced by the ``ideal'' location of the ozone layer, just above the upper haze. However, this relies on the assumption that the O$_3$ distribution is constant with latitude all around Venus, which should be regarded with extreme caution. In the range 200--240~nm, the intense absorption system $\tilde{C}^1B_2$--$\tilde{X}^1A_1$ of SO$_2$ \citep{Freeman:1984} dominates the spectrum. 

Although \ion{O}{i} becomes the main component in the upper atmosphere above 150~km, the column density is too low at these altitudes and the 131-nm line too thin to produce significant absorption along the line of sight. The upper atmosphere of a planet similar to Venus, whose lower atmosphere is dominated by CO$_2$, is too cold ($\sim 200$~K), and the scale height too small, to give rise to detectable signatures from exospheric atoms. This contrasts with the warmer ($\sim1\,000$~K) upper atmosphere of the Earth, where high-altitude atomic species should be easier to detect than on Venus. In fact, on Venus infrared emissions from CO$_2$ molecules (quenched at low altitudes) are the main source of atmospheric cooling above the mesosphere. It would be difficult to detect atmospheric signatures from an Earth-size exoplanet with a small thermospheric scale height. If the precision of the measurements is high enough, a lack of detection could then be attributed to the presence of an efficient upper atmosphere cooler (very likely CO$_2$). In such a planet, the greenhouse effect would drastically change the surface temperature and the criteria used to decide if it lies in the habitable zone.

The only spectroscopic data against which we could validate the present model are the near-infrared observations of the 2004 Venus transit by \citet{Hedelt:2011}. These authors observed the transit with an instrument designed to measure the solar photosphere, the Vacuum Tower Telescope (VTT) in Tenerife. Out of spatially resolved, dispersed images, \citet{Hedelt:2011} obtain high-resolution ($R\sim200\,000$) transmission spectra over narrow wavelength ranges. A direct comparison with our model is uneasy because the observations are spatially resolved over only a small fraction of the solar disc and of Venus' limb. We use a correction factor equal to the ratio of the solar surface covered by the spectrograph slit to the surface of the whole Sun ($\sim87\,000$), allowing us to roughly compare our model to \citet{Hedelt:2011}'s observations in the spectral region 1\,597.5--1\,598.3~nm (see Fig.~\ref{fig:hedelt11}). Given the uncertainties in the data and in the rescaling factor used, there is a fair agreement between our model and their data.

Transmission spectroscopy is a powerful tool to probe the atmospheres of transiting exoplanets. Current challenges involve identifying the scatterers responsible for hot jupiter spectra \citep{Sing:2011a}, understanding the ``flat'' spectra of super-earths \citep{Bean:2011}, and determining the observational differences between habitable and hostile worlds. Mie scattering by clouds or hazes is likely to play a central role in these problems.

The June 2012 transit will be a unique occasion to collect non-spatially resolved transmission spectra of a planet that was long-believed to be the Earth's twin sister. Enshrouded by clouds and hazes, Venus still hides numerous mysteries in spite of its proximity, such as the nature of ``absorber X'' detected by the \emph{Vega} entry probes \citep{Bertaux:1996}. A spatially unresolved and multi-wavelength coverage of the spectral region impacted by Mie scattering, sulfur dioxide, carbon dioxide, and absorber X could give access to these components, validating and bringing new constraints to the model presented here. 

The \emph{Tess} and \emph{Plato} missions from NASA and ESA, respectively, are designed to detect ``Earth's twins'' transiting bright stars before and beyond the end of the decade. Their transmission spectra could be obtained with the \emph{James Webb Space Telescope}. The next transit of Venus will provide key diagnostic tools to discriminate between Earth-like and Venus-like atmospheres of exoplanets transiting their stars.

\acknowledgement
We thank the referee W.~Traub for a prompt and thoughtful report, J.~Pasachoff for discussions around this work, and V.~Wilquet for inputs on the particle size distribution of the upper haze.
 
\bibliographystyle{aa}
\bibliography{aa18400.bib} 

\begin{thebibliography}{41}
\expandafter\ifx\csname natexlab\endcsname\relax\def\natexlab#1{#1}\fi

\bibitem[{Bean {et~al.}(2011)Bean, D{\'e}sert, Kabath, Stalder, Seager,
  Kempton, Berta, Homeier, Walsh, \& Seifahrt}]{Bean:2011}
Bean, J.~L., D{\'e}sert, J.-M., Kabath, P., {et~al.} 2011, arXiv, astro-ph.EP

\bibitem[{Bean {et~al.}(2010)Bean, Kempton, \& Homeier}]{Bean:2010}
Bean, J.~L., Kempton, E. M.-R., \& Homeier, D. 2010, Nature, 468, 669

\bibitem[{Bertaux {et~al.}(1996)Bertaux, Widemann, Hauchecorne, Moroz, \&
  Ekonomov}]{Bertaux:1996}
Bertaux, J.-L., Widemann, T., Hauchecorne, A., Moroz, V.~I., \& Ekonomov, A.~P.
  1996, Journal of Geophysical Research, 101, 12709

\bibitem[{Charbonneau {et~al.}(2009)Charbonneau, Berta, Irwin, Burke, Nutzman,
  Buchhave, Lovis, Bonfils, Latham, Udry, Murray-Clay, Holman, Falco, Winn,
  Queloz, Pepe, Mayor, Delfosse, \& Forveille}]{Charbonneau:2009}
Charbonneau, D., Berta, Z.~K., Irwin, J., {et~al.} 2009, Nature, 462, 891

\bibitem[{Charbonneau {et~al.}(2002)Charbonneau, Brown, Noyes, \&
  Gilliland}]{Charbonneau:2002}
Charbonneau, D., Brown, T.~M., Noyes, R.~W., \& Gilliland, R.~L. 2002, ApJ,
  568, 377

\bibitem[{Cotton {et~al.}(2011)Cotton, Bailey, Crisp, \& Meadows}]{Cotton:2011}
Cotton, D.~V., Bailey, J., Crisp, D., \& Meadows, V.~S. 2011, arXiv,
  astro-ph.EP

\bibitem[{de~Bergh {et~al.}(2006)de~Bergh, Moroz, Taylor, Crisp, B{\'e}zard, \&
  Zasova}]{Bergh:2006}
de~Bergh, C., Moroz, V.~I., Taylor, F.~W., {et~al.} 2006, Planetary and Space
  Science, 54, 1389

\bibitem[{D{\'e}sert {et~al.}(2011)D{\'e}sert, Bean, Kempton, Berta,
  Charbonneau, Irwin, Fortney, Burke, \& Nutzman}]{Desert:2011a}
D{\'e}sert, J.-M., Bean, J., Kempton, E. M.-R., {et~al.} 2011, arXiv,
  astro-ph.EP

\bibitem[{Ehrenreich {et~al.}(2006)Ehrenreich, Tinetti, {Lecavelier des
  Etangs}, Vidal-Madjar, \& Selsis}]{Ehrenreich:2006b}
Ehrenreich, D., Tinetti, G., {Lecavelier des Etangs}, A., Vidal-Madjar, A., \&
  Selsis, F. 2006, A{\&}A, 448, 379

\bibitem[{{Esposito} {et~al.}(1997){Esposito}, {Bertaux}, {Krasnopolsky},
  {Moroz}, \& {Zasova}}]{Esposito:1997}
{Esposito}, L.~W., {Bertaux}, J.-L., {Krasnopolsky}, V., {Moroz}, V.~I., \&
  {Zasova}, L.~V. 1997, in Venus II: Geology, Geophysics, Atmosphere, and Solar
  Wind Environment, ed. {S.~W.~Bougher, D.~M.~Hunten, \& R.~J.~Phillips},
  415--+

\bibitem[{Freeman {et~al.}(1984)Freeman, Yoshino, Esmond, \&
  Parkinson}]{Freeman:1984}
Freeman, D.~E., Yoshino, K., Esmond, J.~R., \& Parkinson, W.~H. 1984, Planetary
  and Space Science (ISSN 0032-0633), 32, 1125

\bibitem[{Gronoff {et~al.}(2008)Gronoff, Lilensten, Simon, Barth{\'e}lemy,
  Leblanc, \& Dutuit}]{Gronoff:2008}
Gronoff, G., Lilensten, J., Simon, C., {et~al.} 2008, A{\&}A, 482, 1015

\bibitem[{Hedelt {et~al.}(2011)Hedelt, Alonso, Brown, Vera, Rauer, Schleicher,
  Schmidt, Schreier, \& Titz}]{Hedelt:2011}
Hedelt, P., Alonso, R., Brown, T., {et~al.} 2011, arXiv, astro-ph.EP

\bibitem[{Hedin {et~al.}(1983)Hedin, Niemann, Kasprzak, \& Seiff}]{Hedin:1983}
Hedin, A.~E., Niemann, H.~B., Kasprzak, W.~T., \& Seiff, A. 1983, J. Geophys.
  Res., 88, 73

\bibitem[{Hummel {et~al.}(1988)Hummel, Shettle, \& Longtin}]{Hummel:1988}
Hummel, J.~R., Shettle, E.~P., \& Longtin, D.~R. 1988, A new background
  stratospheric aerosol model for use in atmospheric radiation models,
  AFGL-TR-88-0166, Air Force Geophys.\ Lab., Hanscom Air Force Base, Mass.

\bibitem[{Ignatiev {et~al.}(2009)Ignatiev, Titov, Piccioni, Drossart,
  Markiewicz, Cottini, Roatsch, Almeida, \& Manoel}]{Ignatiev:2009}
Ignatiev, N.~I., Titov, D.~V., Piccioni, G., {et~al.} 2009, J. Geophys. Res.,
  114

\bibitem[{Kaltenegger \& Traub(2009)}]{Kaltenegger:2009a}
Kaltenegger, L. \& Traub, W.~A. 2009, ApJ, 698, 519

\bibitem[{Knollenberg \& Hunten(1980)}]{Knollenberg:1980}
Knollenberg, R.~G. \& Hunten, D.~M. 1980, J. Geophys. Res., 85, 8039

\bibitem[{Knutson {et~al.}(2011)Knutson, Madhusudhan, Cowan, Christiansen,
  Agol, Deming, D\'esert, Charbonneau, Henry, Homeier, Langton, Laughlin, \&
  Seager}]{Knutson:2011}
Knutson, H.~A., Madhusudhan, N., Cowan, N.~B., {et~al.} 2011, arXiv,
  astro-ph.EP

\bibitem[{{Lecavelier des Etangs} {et~al.}(2008{\natexlab{a}}){Lecavelier des
  Etangs}, Pont, Vidal-Madjar, \& Sing}]{Etangs:2008a}
{Lecavelier des Etangs}, A., Pont, F., Vidal-Madjar, A., \& Sing, D.
  2008{\natexlab{a}}, A{\&}A, 481, L83

\bibitem[{{Lecavelier des Etangs} {et~al.}(2008{\natexlab{b}}){Lecavelier des
  Etangs}, Vidal-Madjar, D{\'e}sert, \& Sing}]{Etangs:2008b}
{Lecavelier des Etangs}, A., Vidal-Madjar, A., D{\'e}sert, J.-M., \& Sing, D.
  2008{\natexlab{b}}, Astronomy and Astrophysics, 485, 865

\bibitem[{L{\'e}ger {et~al.}(2011)L{\'e}ger, Grasset, Fegley, Codron, Albarede,
  Barge, Barnes, Cance, Carpy, Catalano, Cavarroc, Demangeon, Ferraz-Mello,
  Gabor, Griessmeier, Leibacher, Libourel, Maurin, Raymond, Rouan, Samuel,
  Schaefer, Schneider, Schuller, Selsis, \& Sotin}]{Leger:2011}
L{\'e}ger, A., Grasset, O., Fegley, B., {et~al.} 2011, arXiv, astro-ph.EP

\bibitem[{L{\'e}ger {et~al.}(2004)L{\'e}ger, Selsis, Sotin, Guillot, Despois,
  Mawet, Ollivier, Lab{\`e}que, Valette, Brachet, Chazelas, \&
  Lammer}]{Leger:2004}
L{\'e}ger, A., Selsis, F., Sotin, C., {et~al.} 2004, Icarus, 169, 499

\bibitem[{Marov(2005)}]{Marov:2005}
Marov, M.~Y. 2005, Transits of Venus: New Views of the Solar System and Galaxy,
  209

\bibitem[{Montmessin {et~al.}(2011)Montmessin, Bertaux, Lef{\`e}vre, Marcq,
  Belyaev, G{\'e}rard, Korablev, Fedorova, Sarago, \&
  Vandaele}]{Montmessin:2011}
Montmessin, F., Bertaux, J.-L., Lef{\`e}vre, F., {et~al.} 2011, Icarus, 216, 82

\bibitem[{Pasachoff {et~al.}(2011)Pasachoff, Schneider, \&
  Widemann}]{Pasachoff:2011}
Pasachoff, J.~M., Schneider, G., \& Widemann, T. 2011, The Astronomical
  Journal, 141, 112

\bibitem[{P{\"a}tzold {et~al.}(2007)P{\"a}tzold, H{\"a}usler, Bird, Tellmann,
  Mattei, Asmar, Dehant, Eidel, Imamura, Simpson, \& Tyler}]{Patzold:2007}
P{\"a}tzold, M., H{\"a}usler, B., Bird, M.~K., {et~al.} 2007, Nature, 450, 657

\bibitem[{Pont {et~al.}(2008)Pont, Knutson, Gilliland, Moutou, \&
  Charbonneau}]{Pont:2008}
Pont, F., Knutson, H., Gilliland, R.~L., Moutou, C., \& Charbonneau, D. 2008,
  Monthly Notices of the Royal Astronomical Society, 385, 109, (c) Journal
  compilation {\copyright} 2008 RAS

\bibitem[{Rothman {et~al.}(2005)Rothman, Jacquemart, Barbe, Benner, Birk,
  Brown, Carleer, Chackerian, Chance, Coudert, Dana, Devi, Flaud, Gamache,
  Goldman, Hartmann, Jucks, Maki, Mandin, Massie, Orphal, Perrin, Rinsland,
  Smith, Tennyson, Tolchenov, Toth, Auwera, Varanasi, \& Wagner}]{Rothman:2005}
Rothman, L.~S., Jacquemart, D., Barbe, A., {et~al.} 2005, Journal of
  Quantitative Spectroscopy and Radiative Transfer, 96, 139

\bibitem[{Rothman {et~al.}(1998)Rothman, Rinsland, Goldman, Massie, Edwards,
  Flaud, Perrin, Camy-Peyret, Dana, Mandin, Schroeder, McCann, Gamache,
  Wattson, Yoshino, Chance, Jucks, Brown, Nemtchinov, \&
  Varanasi}]{Rothman:1998}
Rothman, L.~S., Rinsland, C.~P., Goldman, A., {et~al.} 1998, JQSRT, 60, 665

\bibitem[{Schneider {et~al.}(2006)Schneider, Pasachoff, \&
  Willson}]{Schneider:2006a}
Schneider, G., Pasachoff, J.~M., \& Willson, R.~C. 2006, ApJ, 641, 565

\bibitem[{Sing {et~al.}(2011)Sing, Pont, Aigrain, Charbonneau, D\'esert,
  Gibson, Gilliland, Hayek, Henry, Knutson, {Lecavelier des Etangs}, Mazeh, \&
  Tal-Or}]{Sing:2011a}
Sing, D.~K., Pont, F., Aigrain, S., {et~al.} 2011, arXiv, astro-ph.EP

\bibitem[{Sneep \& Ubachs(2005)}]{Sneep:2005}
Sneep, M. \& Ubachs, W. 2005, Journal of Quantitative Spectroscopy and
  Radiative Transfer, 92, 293

\bibitem[{Snellen {et~al.}(2010)Snellen, de~Kok, de~Mooij, \&
  Albrecht}]{Snellen:2010}
Snellen, I. A.~G., de~Kok, R.~J., de~Mooij, E. J.~W., \& Albrecht, S. 2010,
  Nature, 465, 1049

\bibitem[{Stevenson {et~al.}(2010)Stevenson, Harrington, Nymeyer, Madhusudhan,
  Seager, Bowman, Hardy, Deming, Rauscher, \& Lust}]{Stevenson:2010}
Stevenson, K.~B., Harrington, J., Nymeyer, S., {et~al.} 2010, Nature, 464, 1161

\bibitem[{Tanga {et~al.}(2011)Tanga, Widemann, Sicardy, Pasachoff, Arnaud,
  Comolli, Rondi, \& S\"utterlin}]{Tanga:2011}
Tanga, P., Widemann, T., Sicardy, B., {et~al.} 2011, Icarus, accepted

\bibitem[{Titov {et~al.}(2009)Titov, Svedhem, Taylor, Barabash, Bertaux,
  Drossart, Formisano, H{\"a}usler, Korablev, Markiewicz, Nevejans,
  P{\"a}tzold, Piccioni, Sauvaud, Zhang, Witasse, Gerard, Fedorov,
  Sanchez-Lavega, Helbert, \& Hoofs}]{Titov:2009}
Titov, D.~V., Svedhem, H., Taylor, F.~W., {et~al.} 2009, Solar System Research,
  43, 185

\bibitem[{Vidal-Madjar {et~al.}(2010)Vidal-Madjar, Arnold, Ehrenreich, Ferlet,
  {Lecavelier des Etangs}, Bouchy, Segransan, Boisse, H{\'e}brard, Moutou,
  D{\'e}sert, Sing, Cabanac, Nitschelm, Bonfils, Delfosse, Desort, Diaz,
  Eggenberger, Forveille, Lagrange, Lovis, Pepe, Perrier, Pont, Santos, \&
  Udry}]{Vidal-Madjar:2010}
Vidal-Madjar, A., Arnold, L., Ehrenreich, D., {et~al.} 2010, Astronomy and
  Astrophysics, 523, 57

\bibitem[{Vidal-Madjar {et~al.}(2003)Vidal-Madjar, {Lecavelier des Etangs},
  D{\'e}sert, Ballester, Ferlet, H{\'e}brard, \& Mayor}]{Vidal-Madjar:2003}
Vidal-Madjar, A., {Lecavelier des Etangs}, A., D{\'e}sert, J.-M., {et~al.}
  2003, Nature, 422, 143

\bibitem[{Vidal-Madjar {et~al.}(2011)Vidal-Madjar, Sing, {Lecavelier des
  Etangs}, Ferlet, D{\'e}sert, H{\'e}brard, Boisse, Ehrenreich, \&
  Moutou}]{Vidal-Madjar:2011}
Vidal-Madjar, A., Sing, D.~K., {Lecavelier des Etangs}, A., {et~al.} 2011,
  Astronomy {\&} Astrophysics, 527, 110

\bibitem[{Wilquet {et~al.}(2009)Wilquet, Fedorova, Montmessin, Drummond,
  Mahieux, Vandaele, Villard, Korablev, \& Bertaux}]{Wilquet:2009}
Wilquet, V., Fedorova, A., Montmessin, F., {et~al.} 2009, J. Geophys. Res., 114

\end{thebibliography}

\appendix

\section{On-line material}
\setcounter{figure}{2}
\begin{figure}[!h]
\resizebox{\columnwidth}{!}{\includegraphics{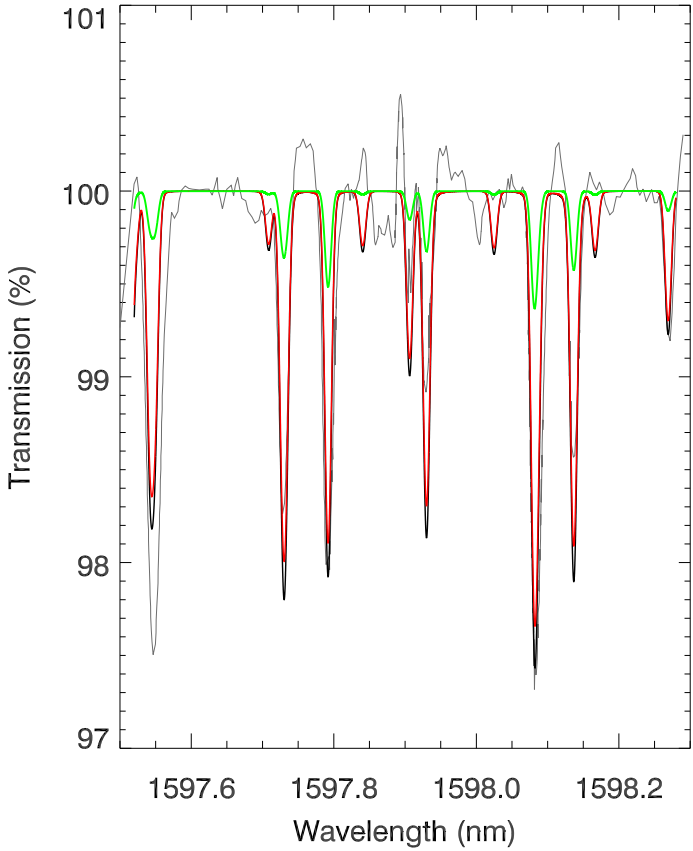}}
\caption{\label{fig:hedelt11} Upper-haze free (black line), mode-1, and modes-1+2 upper-haze models (red and green lines, respectively) compared to the high-resolution limb transmission spectrum observed by \citet{Hedelt:2011} (dark grey line). All spectral lines are CO$_2$ transitions, including isotopes $^{12}$C$^{16}$O$_2$, $^{13}$C$^{16}$O$_2$, and $^{16}$O$^{12}$C$^{18}$O.}
\end{figure} 

\begin{table}[!h]
\begin{center}
\begin{tabular}{*{8}{c}}
\hline\hline
Wavelength & \multicolumn{3}{c}{Absorption} && \multicolumn{3}{c}{Effective height} \\
\cline{2-4}\cline{6-8}
           & no haze & mode-1 haze & modes-1+2 haze && no haze & mode-1 haze & modes-1+2 haze \\ 
(nm)       & (ppm)   & (ppm)       & (ppm)          && (km)    & (km)        & (km) \\
\hline
100.00000 & 25.245750 & 25.112910 & 23.1596701 && 49.668261 & 49.274121 & 49.27412 \\
101.00000 & 25.069860 & 24.937020 & 22.9837801 && 49.115681 & 48.721501 & 48.72150 \\
102.00000 & 26.222370 & 26.089530 & 24.1362901 && 52.735541 & 52.341591 & 52.34159 \\
\ldots    & \ldots    & \ldots    & \ldots     && \ldots    & \ldots    & \ldots   \\
\hline
\end{tabular}
\caption{\label{tab:spectra}Transit spectra of Venus from Figs.~\ref{fig:spectrum}b (absorption) and~\ref{fig:spectrum}c (effective height). The `no haze', `mode-1 haze', and `modes-1+2 haze' columns correspond to the black, red, and green curves in these figures, respectively. This Table is only  available in electronic form at the CDS via anonymous ftp to cdsarc.u-strasbg.fr (130.79.128.5) or via \texttt{http://cdsweb.u-strasbg.fr/cgi-bin/qcat?J/A+A/}.}
\end{center}
\end{table}
 
\end{document}